\documentclass{cta-author}

{}
{}
{}

\usepackage{epsf}
\usepackage{graphicx}
\usepackage{floatflt}
\usepackage{color}
\usepackage{comment}
\usepackage[utf8]{inputenc}
\usepackage{amssymb}
\usepackage{amsmath}
\usepackage{subcaption}
\usepackage{algorithm}
\usepackage{algpseudocode}
\usepackage{setspace}
\usepackage{ulem}
\usepackage{url}
\usepackage{supertabular,booktabs}

\floatname{algorithm}{Procedure}

\begin{document}

\supertitle{Submission Template for IET Research Journal Papers}

\title{Hybrid data-driven physics model-based framework for enhanced cyber-physical smart grid security}

\author{\au{Cody Ruben$^{1\corr}$}, \au{Surya Dhulipala$^{1}$}, \au{Keerthiraj Nagaraj$^{1}$}, \au{Sheng Zou$^{1}$}, \au{Allen Starke$^{1}$}, \au{Arturo Bretas$^{1}$}, \au{Alina Zare$^{1}$}, \au{Janise McNair$^{1}$}}

\address{\add{1}{Electrical and Computer Engineering, University of Florida, 1064 Center Dr, Gainesville, FL, USA}
\email{cruben31@ufl.edu}}

\begin{abstract}
This paper presents a hybrid data-driven physics model-based framework for real time monitoring in smart grids. As the power grid transitions to the use of smart grid technology, it's real time monitoring becomes more vulnerable to cyber attacks like false data injections (FDI). Although smart grids cyber-physical security has an extensive scope, this paper focuses on FDI attacks, which are modeled as bad data. State of the art strategies for FDI detection in real time monitoring rely on physics model-based weighted least squares state estimation solution and statistical tests. This strategy is inherently vulnerable by the linear approximation and the companion statistical modeling error, which means it can be exploited by a coordinated FDI attack. In order to enhance the robustness of FDI detection, this paper presents a framework which explores the use of data-driven anomaly detection methods in conjunction with physics model-based bad data detection via data fusion.  Multiple anomaly detection methods working at both the system level and distributed local detection level are fused. The fusion takes into consideration the confidence of the various anomaly detection methods to provide the best overall detection results. Validation considers tests on the IEEE 118 bus system. 
\end{abstract}

\maketitle
This paper is a preprint of a paper submitted to IET Smart Grid. If accepted, the
copy of record will be available at the IET Digital Library.
\section{Introduction}\label{sec:intro}

The future power grid, or Smart Grid (SG), will integrate control, communication and computation aiming to achieve stability, efficiency and robustness of the physical processes on the system.  These advancements bring many challenges to the SG and therefore have drawn much attention from academia, industry and government due to the great impact they will have on society, economics and the environment. While a great amount of research has been done towards these objectives, science and technology related to the cyber-physical security of SGs are still immature. Additionally, much of the critical infrastructure is currently transitioning towards the paradigm of SGs by increasing the dependency of control of physical processes on communication networks, thus becoming exposed to cyber-threats~\cite{faragSGSecurity2014}. 


\color{black}
Guaranteeing the reliable operation of power grids is crucial for today's society and it is done through real-time power system monitoring. Currently, real-time monitoring is done through a process called power system state estimation (PSSE) \cite{monticelli1999state}, which provides relevant information on the power grid current operating point based on the measurements throughout the system. These measurements are commonly transmitted to a Supervisory Control and Data Acquisition (SCADA) system, which implements centralized monitoring and control for the electrical grid, where PSSE is performed. One important feature of PSSE is its error processing capability. Measurements that are clearly inconsistent are discarded in the pre-filtering step, which precedes state estimation step. Following state estimation using pre-filtered data, a post-processing step called bad data analysis is performed. This step aims at detecting bad data or gross errors (GE), assuming they are statistically large errors. 

The established procedure to determine the current operating point for the power grid has been based on iterative numerical linearization around the incumbent solution, which is determined easily by least squares approaches to find solutions within the convex hull. There is always an inherent error in this approximation procedure that is difficult to quantify, but it has been sufficiently small for the current requirements of dispatching, detection of failures, and reliability studies. 
This linear approximation and the companion statistical modeling error approach alone are not compatible to the new demands of cyber-physical security because its sensitivity and specificity is theoretically lower bounded by the level and dynamics of the approximation error, which can be easily explored by coordinated cyber-attacks.

The increasing dependence on digital monitoring and control of power systems raises concerns with respect to cybersecurity. One common type of cyber-threat is false data injection (FDI) attack, where an attack aims to disrupt the operation of the power grid by modifying a subset of measurement values. While bad data analysis is capable of detecting many instances of Gross Errors via tests such as the Chi Squared Test, largest normalized residual \cite{handschin1975} or innovation-based \cite{bretas2011} approaches, intelligent cyber-attacks may be engineered to be difficult to detect \cite{liu2009}, considering the implicit constraints of physics model-based solutions. 

Methods devised to tackle FDI attacks include Generalized Likelihood Ratio Detector with L-1 Norm Regularization \cite{kosut2010}, a scheme for protecting a selected set of measurements and verifying the values of a set of state variables independently \cite{bobba2010}, and the estimation of the normalized composed measurement error for detection of malicious data attacks \cite{bretas2017}. These solutions all consider a quasi-static physics measurement model, however the power grid is a time-varying system with loads and generation constantly changing. Thus grid temporal characteristics are not fully explored for FDI detection.  

Considering environment temporal characteristics, machine learning based solutions have also been explored for anomaly detection. The Correlation-based Detection (CorrDet) algorithm \cite{ho2000correlation} was introduced for landmine detection. The Reed-Xiaoli (RX) Detector \cite{chang2002anomaly} was introduced for target detection of remote sensing images. Both methods rely on the  approach where an incoming sample is classified as abnormal if its squared Mahalanobis distance with respect to a background statistic is above some threshold.

\begin{table}
    \centering
    \label{tab:math_notations}
    \begin{tabular}{|p{0.15\columnwidth}|p{0.75\columnwidth}|}
        \toprule
        \textbf{Notation} & \textbf{Explanation}\\
        \midrule
        $J(\mathbf{x})$ & WLS cost function - $\mathbb{R}^{1 \times 1}$  \\ 
        
        $d$ & number of measurements - $\mathbb{R}^{1 \times 1}$ \\
        
        $\mathbf{P}$ & projection matrix - $\mathbb{R}^{d \times d}$ \\
        
        $\alpha_{\chi}$ & significance level - $\mathbb{R}^{1 \times 1}$ \\
        
        $\mathbf{z}$ & vector of measurements - $\mathbb{R}^{1 \times d}$ \\
        $\mathbf{z}^*$ & operating measurement vector - $\mathbb{R}^{1 \times d}$ \\
        $\mathbf{\hat{z}}$ & estimated measurement vector - $\mathbb{R}^{1 \times d}$ \\
        
        
        
        
        $\mathbf{Z}$ & training set - $\mathbb{R}^{d \times K_{1}}$  \\
        
        $\mathbf{\hat{Z}}$ & testing set - $\mathbb{R}^{d \times K_{2}}$  \\
        
        $K_{1}$ & number of samples in training set - $\mathbb{R}^{1 \times 1}$ \\
        
        $K_{2}$ & number of samples in testing set - $\mathbb{R}^{1 \times 1}$ \\
        
        $\mathbf{x}$ & vector of state variables - $\mathbb{R}^{1 \times N}$ \\
        
        $\mathbf{x^*}$ & operating state vector - $\mathbb{R}^{1 \times N}$ \\
        
        $\mathbf{\hat{x}}$ & estimated state vector - $\mathbb{R}^{1 \times N}$ \\
        
        
        
        $N$ & number of states - $\mathbb{R}^{1 \times 1}$ \\
        $\mathbf{\Sigma}_{SE}$  & covariance of measurement vector for state estimator- $\mathbb{R}^{d \times d}$\\
        
        $\mathbf{\Sigma}$  & covariance of measurement vector - $\mathbb{R}^{d \times d}$\\
        
        $\mathbf{e}$  &  vector of measurement error - $\mathbb{R}^{1 \times d}$ \\
        
        $e_{D}$ & detectable component of error - $\mathbb{R}^{1 \times 1}$ \\
        
        $e_{U}$ & undetectable component of error - $\mathbb{R}^{1 \times 1}$ \\
        
        $II$ & Innovation Index of measurement - $\mathbb{R}^{1 \times 1}$ \\
        
        $\sigma$ & standard deviation of e - $\mathbb{R}^{1 \times 1}$ \\
        
        $\mathbf{h(x)}$ & vector of measurement estimates - $\mathbb{R}^{d \times d}$ \\
        
        $\mathbf{H}$ & Jacobian matrix - $\mathbb{R}^{d \times N}$ \\
        
        $\mathcal B$ & set of buses $j$ $\ni$ $Y_{ij}\neq 0$ \\

        $\chi_{d,p}^{2}$ & Chi-squared value - $\mathbb{R}^{1 \times 1}$  \\
        
        $\mathbf{r}$ & vector of measurement residuals - $\mathbb{R}^{1 \times d}$  \\
        
        $CME$ & Composed Measurement Error - $\mathbb{R}^{1 \times 1}$  \\
        
        $W_t$  & driving noise of O-U process \\
        
        $\beta$ & decay-rate of O-U process \\
        
        $\sigma^2_{n}$  & variance of noise of O-U process \\
        
        $\mu_{o-u}(t)$ & long term mean of O-U process \\

        $\delta^{CD}$ & squared Mahalanobis distance value - $\mathbb{R}^{1 \times 1}$  \\
        
        $\mathbf{\mu}$ & mean of measurement vector - $\mathbb{R}^{1 \times d}$  \\
        
        $\tau$ & threshold value to classify abnormal samples - $\mathbb{R}^{1 \times 1}$ \\
        $\mathbf{T}$ & set of $\tau$  ($\mathbb{R}^{1 \times M}$) \\
        
        
        
        $l$ & label of a sample - $\mathbb{R}^{1 \times 1}$  \\
        
        $\alpha$ &  weight value - $\mathbb{R}^{1 \times 1}$  \\
        
        $M$ & number of buses - $\mathbb{R}^{1 \times 1}$  \\
        
        $\mathbf{Y}$ & label for training samples - $\mathbb{R}^{1 \times K_{1}}$  \\
        
        $\mathbf{\hat{Y}}$ & label for testing samples -$\mathbb{R}^{1 \times K_{2}}$\\
        
        $\phi_m$ & symbol for m-th Local CorrDet detector \\
        
        $\Phi_E$ & symbol for Ensemble CorrDet detector \\
        
        $\Phi_R$ & symbol for CorrDet detector \\
        
        $\mathbf{\delta}_{Z,m}$ &  $\mathbf{\delta}^{CD}$ of all training samples with respect to $\phi_m$ - $\mathbb{R}^{1 \times K_{1}}$ \\
        
        $\mathbf{\delta}_{z_{\hat{k}}}$ & $\mathbf{\delta}^{CD}$ of k-th testing sample with respect to $\Phi_E$ - $\mathbb{R}^{1 \times M}$ \\
        
        $\mathbf{\Psi}_{ECD}$ & overall decision scores from Ensemble CorrDet detector - $\mathbb{R}^{1 \times K_{2}}$ \\
        
        $\mathbf{\Psi}_{SE}$ & decision scores from state estimator technique - $\mathbb{R}^{1 \times K_{2}}$ \\
        
        $\mathbf{\Psi}_{fusion}$ & fusion decision scores for hybrid data-driven physics model-based framework - $\mathbb{R}^{1 \times K_{2}}$ \\
        \hline
        
    \end{tabular}%
\end{table}

Some artificial  intelligence-related  detection  methods  have  been  put forward in recent years, which are mainly based on neural network, deep learning and fuzzy clustering ~\cite{realtime-detection,machinelearningdatainjection}. In ~\cite{DBN}, the authors focus on the detection of false data  injection  attacks  in  smart grids using a deep belief network-based (DBN) attack detection method with unsupervised learning methods to provide the initial weights. The authors of ~\cite{ANNFDIA}, overcome  the  above-mentioned  problems with Artificial Neural Network (ANN) based approach which identifies the FDI by tracking the measurement data. The  uniqueness  of  the  neural  network  method  is  simple  infrastructure but uneasy in the parameter adjustment. Numerous tests should be utilized to train the network model. A deep learning method originates from the neural network,  which  can  solve  the  overfitting  problems  well  but  the training method is more complex ~\cite{reviewFDIA}. Unsupervised   learning   is   performed   from   the   bottom   of   the   restricted   Boltzmann   machine   to provide   initial   weights   for   the   network.   The   backpropagation  algorithm  propagates  the  error  from  top  to  bottom  and  fine-tunes  the  model  parameters.

Considering hybrid data-driven physics model-based solutions for FDI detection, literature review will present seldom contributions. It is clear that analytic quasi-static model-based solutions can represent spatial characteristics of the environment, while data-driven solutions can explore temporal characteristics which are inherent to the grid operation. Thus they are complementary solutions. Considering such rationale, in \cite{trevizan2019}, a previous work of the authors, an extended Chi-squared test using information from PSSE and a data driven CorrDet algorithm \cite{ho2000correlation} is presented.  

\begin{figure*}[ht]
\begin{center}
\includegraphics[width=1.9\columnwidth]{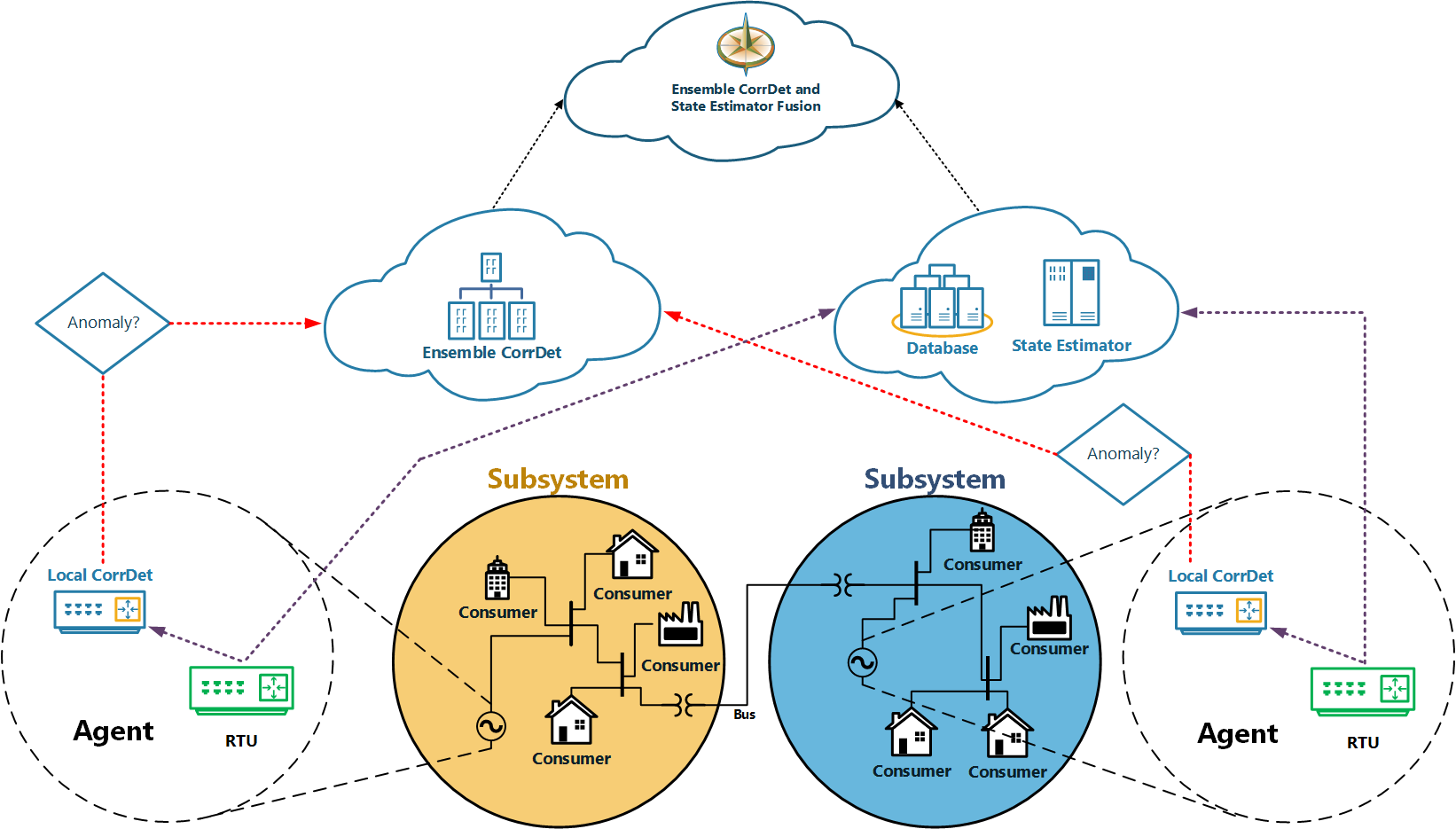}
\caption{Distributed multi-agent architecture}
\label{fig:dist-sub}
\end{center}
\end{figure*}

In this work, a hybrid data-driven physics model-based framework for FDI detection on system real-time monitoring is presented. Figure \ref{fig:dist-sub} illustrates the presented framework, which explores both temporal and spatial characteristics of the environment in a distributed multi-agent architecture. An Ensemble CorrDet algorithm is presented to address drifting load scenarios and numerical issues. The CorrDet algorithm is used to learn the background statistics (e.g. mean and covariance for normal samples) over the whole power grid topology, consisting of a series of buses where feature (measurement) values are measured in a single bus or between two buses. Ensemble CorrDet detector can be considered as a set of CorrDet detectors for each local environment. Therefore, Ensemble CorrDet detector learns a series of background statistics, one for each bus. There are several advantages for using Ensemble Corrdet detector compared to Corrdet detector. FDI are sparse, thus learning the background statistics on each bus, instead of the whole power system topology, allows for a more sensitive anomaly detection while embedding local environmental spatial characteristics to the data-driven solution. More specifically, spatially neighboring buses are more highly correlated and easier to be affected by an attack while buses that are further away have lower correlation. Thus, learning a full covariance over all measurements of all buses is unnecessary (nearly sparse covariance), especially when training data is limited. Instead, local, fewer dimensional measurement sets offer a more accurate statistic estimation and a computationally cheaper, more sensitive anomaly detection. Ensemble CorrDet detector is also a scalable solution as even when more buses are added to the grid, we could just add a Local CorrDet detector to that bus and include the result of this local detector in the Ensemble CorrDet algorithm. Second, learning the background statistics on the whole power system topology is usually more challenging than learning on a bus. The latter allows a distributed local environmental model learning on the much smaller dimensional data (usually several measurements) than the former (usually several hundreds of measurements). To avoid the numerical issues due to large dimensions, a much larger training set would be necessary for the CorrDet detector to achieve the same performance, otherwise, estimated background statistics are often ill-posed. Third, Ensemble CorrDet detector is more robust and secure than CorrDet detector, since the classification of the anomaly for Ensemble CorrDet detector is an aggregated decision of a series of local CorrDet detectors, as a \textit{committee}, allowing for a small number of failed local CorrDet detections, while CorrDet detector is a single detector. Lastly, a good feature of Ensemble CorrDet detector is FDI localization. For instance, if an FDI happens on a measurement between two buses and the corresponding two local CorrDet detectors flags the sample as abnormal, it can also be inferred that the FDI is a measurement between the two buses, while CorrDet detector cannot find the location where false data was injected.

In order to create the hybrid between physics model-based and data-driven solutions, a decision level fusion solution is presented. On such, several data-driven and physics-based methods for the FDI attacks detection on SG real-time monitoring are combined producing one output.  The fusion considers the confidence of each anomaly detection method, creating the hybrid method that proves to perform better than any of the individual detection methods.

The contributions of this paper are as follows:
\begin{itemize}
    \item Decision level fusion is used to create a hybrid physics-based data-driven anomaly detection that considers the confidence of individual anomaly detection methods such that different anomalies detected by the individual methods can be aggregated upon fusion;
    \item A data-driven Ensemble CorrDet anomaly detection method is presented which works in a distributed fashion, allowing for more sensitive local detection on the spatial domain;
    \item Using a physics-based bad data detection method considering the Innovation Concept to improve the data driven detection method.
\end{itemize}

The remainder of the paper is organized as follows. In Section \ref{sec:back}, background information of the challenges and existing solutions are presented.  Section \ref{sec:frame} presents the proposed framework, discussing each component and how they are used to build the framework.  
The results of numerical tests used to evaluate the method's performance are shown in Section \ref{sec:case}.  Finally, Section \ref{sec:conc} presents the conclusions of this work.

\section{Background, Challenges, Existing Solutions and their Limitations} \label{sec:back}

\subsection{Smart Grid System}
PSSE plays a crucial role in SG real-time monitoring, as it is responsible for estimating unknown system state variables based mostly on measurements of voltage, power and current. As SG evolution exposes the system to cyber-attacks, it becomes increasingly important to develop countermeasures ~\cite{false1,vul3,on4,reach5,power6,falsedata8}. Most research~\cite{malicious9,detect10,strategic11} in this area focuses on improving bad data detection schemes or improving the security of the communication system. An analysis of the state-of-the-art physics model-based solutions for FDI detection on SGs real-time monitoring will show that these apply residual-based approaches for cyber-attack detection, while ignoring the inherent masked error component ~\cite{bretas2018extension}. Still, as cited in~\cite{cyber2}, a stealthy attack requires the corruption of several measurements. This relates to the fact that a stealthy attack must have the attack vector fitting the measurement model, which is equivalent to shifting the result of the state estimation to a physically possible but wrong solution. Cyber-attacks with such characteristics are hard to discover when applying the classical bad data approaches, which may cause the cyber-attack to remain undetected~\cite{Bretas201943}. 

The authors of ~\cite{dynamicFDIA} proposed an interval forecasting method to predict the possible largest variation bounds of each state variable based on a worst-case analysis based on the forecasting uncertainties of renewable energy sources, and electric loads. Works such as ~\cite{ELM} also use extreme learning machine-based (ELM) one-class-one-network (OCON) and prediction methods to improve the resilience of the power system, exploiting the spatial correlation of power data within subnets.

While power systems are ideally in a steady state, there are constant variations to load and generation.  Current bad data analysis techniques in power systems work in a quasi-steady state, only considering a single snapshot of the system.  In theory, temporal changes on the system could add more information for bad data detection that current techniques are not using.  In \cite{trevizan2019}, the authors begin to make use of this temporal data by developing an extended Chi-squared test that combines the classical PSSE method with a data driven CorrDet algorithm that takes into account past data.  This extended Chi-squared test does not work as well when introduced to more realistic load variation and stealthier FDI attacks.

\subsection{CorrDet and Ensemble CorrDet detectors}

The CorrDet detector is capable of  training a data-driven anomaly detector, however it is not very efficient and may suffer from numerical issues when the number of measurements is large. Therefore, ensemble CorrDet detector is proposed. For the CorrDet detector, only one set of parameters (sample mean, sample covariance matrix and detection threshold) are estimated for the whole power system. These parameters, especially covariance matrix, characterize the correlations among all measurements on all buses. The required number of training samples rapidly grows when the number of measurements increases. Therefore, the detector parameters, especially the covariance matrix, are usually not well-trained, resulting in numerical issues and bad detection performance when the number of training samples is limited. However, learning a full covariance matrix for all measurements is actually not necessary since the covariance matrix is sparse. The reason is that the farther two buses are, the less correlation their measurements will have. In other words, if an FDI happens on one measurement, the most probable affected buses are usually the one or two that the measurement is linked to. Thus, this challenge can be tackled by considering the detection at a smaller spatial scale. More specifically, instead of training the detector parameters at the scale of all measurements of all buses, we can train a set of detectors, which trains measurements associated on each single bus.

Typically, the number of measurements that are linked to each bus is less than ten, while the total number of measurements is several hundred. The reduced dimensionalities for each local CorrDet detector yields more accurate estimation of normal sample means and covariance matrices and finally a more sensitive detection. By looking at which local CorrDet detector(s) reports an anomaly, we can further infer the possible position where the FDI happens.

\section{Hybrid Physics Model-Based Data-Driven Framework} \label{sec:frame}
\subsection{Physics Model} \label{sec:psse}
In modern Energy Management Systems (EMS), the State Estimation (SE) process is the core process for situational awareness of a power system and is used in many EMS applications, including the detection of bad data.  The common approach to SE is using the classical Weighted Least Squares (WLS) method described in \cite{monticelli1999state}.  In this approach, the system is modeled as a set of non-linear algebraic equations based on the physics of the system:

\begin{equation} \mathbf{z=h(x)+e} \label{eq:SE}\end{equation}
where $\mathbf{z}\in\mathbb{R}^{1 \times d}$ is the measurement vector, $\mathbf{x}\in\mathbb{R}^{1 \times N}$ is the vector of state variables, $\mathbf{h}:\mathbb{R}^{1 \times N}\rightarrow\mathbb{R}^{1 \times d}$ is a continuously non-linear differentiable function, and $\mathbf{e}\in\mathbb{R}^{1 \times d}$ is the measurement error vector.  Each measurement error, $e_i$ is assumed to have zero mean, standard deviation $\sigma_i$ and Gaussian distribution.  $d$ is the number of measurements and $N$ is the number of states.

In the classical WLS approach, the best estimate of the state vector in \eqref{eq:SE} is found by minimizing the cost function $\mathbf{J(x)}$:

\begin{equation} 
\label{eq:JSE} 
\mathbf{J(x)=\Vert z-h(x)}\Vert _{\mathbf{\Sigma}^{-1}_{SE}}^{2}=\mathbf{[z-h(x)]'\Sigma}^{-1}_{SE}\mathbf{[z-h(x)]} 
\end{equation}
where $\mathbf{\Sigma}$ is the covariance matrix of the measurements.  In this paper, we consider the standard deviation of each measurement to be 1\% of the measurement magnitude, which has been shown to improve the detection of bad data \cite{BRETAS2015484}. The covariance submatrix for zero injection measurements is calculated as shown in (\ref{eq:zero-inj-1}) and (\ref{eq:zero-inj-2}) . In order to solve this problem, \eqref{eq:SE} is linearized at a certain point $\mathbf{x}^*$ in \eqref{SElin} and the optimal states are found through an iterative process.

\begin{equation} \Delta \mathbf{z=H}\Delta \mathbf{x+e} \label{SElin} \end{equation}
\begin{equation} \sigma(P_{i}) = \sqrt{\sum_{j\in\mathcal B} \sigma^{2}(P_{ij})} \label{eq:zero-inj-1} \end{equation}
\begin{equation} \sigma(Q_{i}) = \sqrt{\sum_{j\in\mathcal B} \sigma^{2}(Q_{ij})} \label{eq:zero-inj-2} \end{equation}
where $\mathbf{H}=\frac{\delta \mathbf{h}}{\delta \mathbf{x}}$ is the Jacobian matrix of $\mathbf{h}$ at the current state estimate $\mathbf{x}^*$, $\Delta \mathbf{z}=\mathbf{z-h(x}^*)=\mathbf{z}-\mathbf{z}^*$ is the correction of the measurement vector and $\Delta \mathbf{x}=\mathbf{x}-\mathbf{x}^*$ is the correction of the state vector. $\mathcal B$ is the set of buses $j$ $\ni$ $Y_{ij}\neq 0$. 
The WLS solution is the projection of $\Delta \mathbf{z}$ onto the Jacobian space by a linear projection matrix $\mathbf{P}$, i.e. $\Delta \mathbf{z=P}\Delta\hat{\mathbf{z}}$.  Letting $\mathbf{r}=\Delta \mathbf{z}-\Delta\hat{\mathbf{z}}$ be the residual vector, the $\mathbf{P}$ matrix that minimizes $\mathbf{J(x)}$ will be orthogonal to the Jacobian range space and to $\mathbf{r}$; $\Delta\hat{\mathbf{z}}=\mathbf{H}\Delta\hat{\mathbf{x}}$.  This is in the form:
\begin{equation} \langle\Delta\hat{\mathbf{z}},\mathbf{r}\rangle=(\mathbf{H}\Delta\hat{\mathbf{x}})'\mathbf{\Sigma}^{-1}_{SE}(\Delta \mathbf{z}-\mathbf{H}\Delta\hat{\mathbf{x}})=0.  \label{WLSsol}\end{equation}

Solving \eqref{WLSsol} for $\Delta\hat{\mathbf{x}}$:
\begin{equation} \Delta\hat{\mathbf{x}}=\mathbf{(H'\Sigma}^{-1}_{SE}H)^{-1}\mathbf{H'\Sigma}^{-1}_{SE}\Delta \mathbf{z}. \label{dx} \end{equation}

At each iteration, a new incumbent solution $\mathbf{x}_{new}^*$ is found and updated following $\mathbf{x}_{new}^*= \mathbf{x}^*+\Delta\hat{\mathbf{x}}$. \eqref{dx} is solved each iteration until $\Delta \hat{\mathbf{x}}$ is sufficiently small to claim convergence of the solution.  Once the SE converges, the Innovation Concept is used for the detection of bad data in the measurement vector $\mathbf{z}$ \cite{bretas2011,bretas2017}.  In the Innovation Concept, the Innovation Index (II) in \eqref{II} is defined as the ratio of the detectable and undetectable components of the measurement error and can be calculated using the projection matrix, $\mathbf{P}$, of the WLS solution \eqref{proj}.

\begin{equation}\label{II}
	II_i=\frac{||e_{Di}||_{\mathbf{\Sigma}^{-1}_{SE}}}{||e_{Ui}||_{\mathbf{\Sigma}^{-1}_{SE}}}=\frac{\sqrt{1-P_{ii}}}{\sqrt{P_{ii}}}.
\end{equation}

\begin{equation}\label{proj}
    \mathbf{P=(H'\Sigma}^{-1}_{SE}\mathbf{H})^{-1}\mathbf{H'\Sigma}^{-1}_{SE}.
\end{equation}

The II and the measurement residual, $r$, are used to calculate the Composed Measurement Error (CME) for each measurement, as shown in \eqref{CME}.

\begin{equation}\label{CME}
	CME_i=r_i\sqrt{1+\frac{1}{II_i^2}}.
\end{equation}

The measurements are considered to be i.i.d, so the statistical Chi-squared test is used, as shown in \eqref{eq:chi-squared}.  

\begin{equation} \label{eq:chi-squared}
\sum_{i=1}^d \left[\frac{CME_i}{\sigma_i}\right]^2 > \chi^2_{\alpha_{\chi},d}
\end{equation}
for $d$ degrees of freedom and significance level $\alpha_{\chi}$. In this paper, the common significance level of 0.05 is used.  
If the sum of normalized CME values is greater than the Chi-squared distribution value, then bad data with $(1-\alpha_{\chi})$ confidence level is detected. We call this statistical test \textit{CME Chi-Square Test} (CMECST).

\subsection{Data-Driven Machine Learning} \label{sec:ddmi}

In addition to the spatial information that is used in physics based-model, data-driven machine learning algorithms (CorrDet and Ensemble CorrDet algorithms) are proposed to take advantage of both temporal and spatial information.  CorrDet algorithm is proposed to estimate the sample statistics globally, while Ensemble CorrDet algorithm combines Local CorrDet estimates, where spatially remote correlations are ignored but the spatially neighboring correlations are reserved.

\subsubsection{CorrDet Anomaly Detection}

\color{black} The machine learning layer of the proposed smart power grid framework uses the knowledge of already verified data to learn the normal state of a properly functioning grid. It is then able to detect any anomalies introduced into the system at any point forward and alerts the cloud layer, as shown in Figure \ref{fig:dist-sub}, to identify the anomaly, isolate it from the remainder of the system and take appropriate action to prevent contamination of the system, with regards to both power distribution in other subsystems, and data assimilation by the machine learning system itself.

A machine learning layer is implemented using the CorrDet Anomaly Detection \cite{ho2002linear, ho2000correlation, chang2002anomaly} algorithm described in equation \eqref{CorrDet-algorithm}, where $\mathbf{z}$ is the new incoming data, $\mathbf{\mu}$ is the mean and $\mathbf{\Sigma}^{-1}$ is the inverse covariance matrix of normal samples. Equation \eqref{CorrDet-algorithm} calculates the squared Mahalanobis distance, $\delta^{CD}(\mathbf{z})$, of a given data $\mathbf{z}$, from the mean, $\mathbf{\mu}$ of the distribution.

\begin{equation} \label{CorrDet-algorithm} \delta^{CD}(\mathbf{z})={(\mathbf{z}-\mathbf{\mu})}^T \mathbf{\Sigma}^{-1} (\mathbf{z}-\mathbf{\mu}). \end{equation}

The anomaly detector is trained with the first $k$ number of incoming samples to generate the $\mathbf{\mu}$ and $\mathbf{\Sigma}^{-1}$. It then accepts new data and uses equation \eqref{CorrDet-algorithm} to determine its squared Mahalanobis distance and compares it to a threshold value $\tau$.


\begin{equation}
  l(\mathbf{z})=\begin{cases}
    1, & \text{if $\delta^{CD}(\mathbf{z}) \geq \tau$}\\
    0, & \text{if $\delta^{CD}(\mathbf{z}) < \tau$}.
  \end{cases}
\end{equation}

If the result is below the threshold, the new data is considered to be normal data (label $l(\mathbf{z})=0$) but if the result is above the threshold, the new data is flagged as an anomaly (label $l(\mathbf{z})=1$). 

In order to select this threshold $\tau$ for Mahalanobis distance, we conduct an experiment to vary $\tau$ as a function of standard deviation ($\sigma_{thr}$) and mean ($\mathbf{\mu}_{thr})$ of Mahalanobis distance values of all the normal samples in training dataset as shown in Equation \ref{eq:threshold}. We use F1-score as the performance metric to choose the value of \textit{$\eta$} of Equation \ref{eq:threshold} that results in the highest F1-score for training data to decide the optimal value of $\tau$.

\begin{equation} \label{eq:threshold} 
\mathbf{\tau} = \mathbf{\mu}_{thr} + \eta * \mathbf{\sigma}_{thr}.
\end{equation}

The $\mathbf{\mu}$ and $\mathbf{\Sigma}^{-1}$ statistics can be updated over time with measurement values from normal test samples to make the model more adaptive and learn the behavior of data over time. 

\subsubsection{Ensemble CorrDet Algorithm} \label{sec:ecd}


The sample data $\mathbf{z}_i$ is usually a set of hundreds of measurements at time $i$, which can be represented as $\mathbf{z}_t = [z_{t1}, z_{t2}, ..., z_{td}]$, where $d$ is the number of measurements and each element $z_{tj}$ denotes the $j$-th measurement value at time $t$. In this work, the dissimilarity comes from the abnormal behaviors of one or several measurements in the power system. 

\begin{figure}[h]
\begin{center}
\includegraphics[height=6cm]{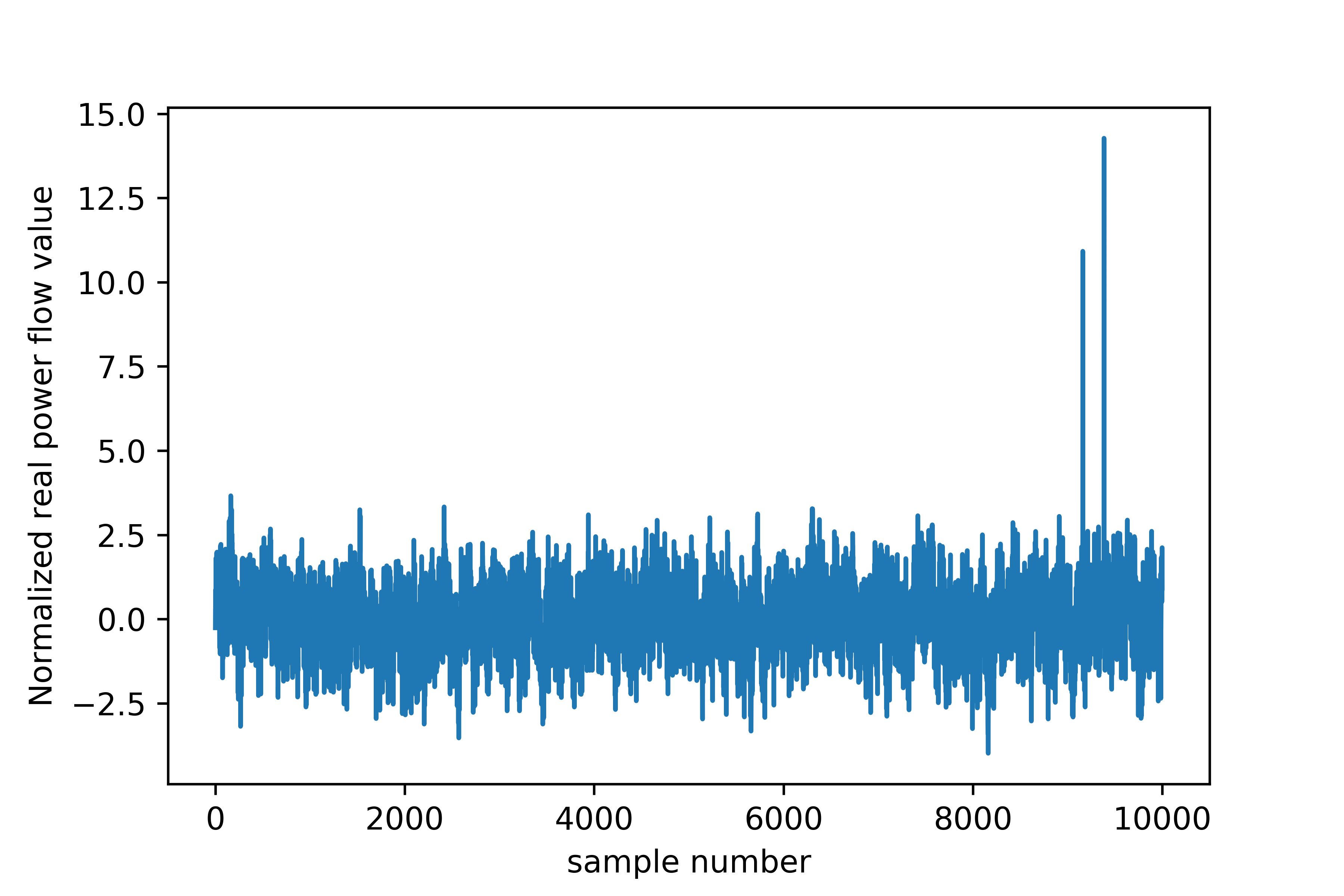}
\caption{Normalized real power flow values from bus 5 to bus 3 for all samples $t = 1$ to $t =10000$}
\label{fig:abnor}
\end{center}
\end{figure}

In other words, an abnormal sample is caused by at least one measurement value that is abnormal.  There are a variety of reasons for a measurement to be abnormal, including an FDI.  For instance, Figure \ref{fig:abnor} shows the normalized real power flow values from bus 5 to bus 3 ($z_{t,340}$) from time $t = 1$ to $t = 10000$. There are 2 measurements $\mathbf{z}_{9161}$ and $\mathbf{z}_{9385}$ in $z_{t,340}$ that are far different than the rest, which can be detected as abnormal measurements. Therefore, the samples $\mathbf{z}_{9161}$ and $\mathbf{z}_{9385}$, containing the abnormal measurement $\mathbf{z}_{9161, 340}$ and $\mathbf{z}_{9385, 340}$, are detected as abnormal samples. 

Since buses are connected via transmission lines, spatially neighboring buses are more highly correlated and easily affected once being attacked while buses that are farther away have less correlation. Thus, learning a full covariance over all measurements of all buses is unnecessary (nearly sparse covariance) when training data is limited. Instead, it is proposed to use local regions with fewer measurements.  These regions have smaller dimension and offer a more accurate normal statistic estimation while being computationally cheaper and provide more sensitive anomaly detection. Based on the CorrDet algorithm, a spatial-temporal, regional CorrDet anomaly detection method is proposed, named \textit{Ensemble CorrDet (ECD)}. 

Ensemble CorrDet detector is defined as a set of Local CorrDet detectors on data samples with a few, spatial neighboring measurements, compared to full measurements in the CorrDet detector. To be more specific, let $\Phi_E$ be the Ensemble CorrDet detector, $\Phi_R$ the CorrDet detector with full dimensionality and $\phi_m$ the Local CorrDet detector with reduced dimensionality where $m = 1:M$ and $M$ is the number of buses. So $\Phi_E = \{\phi_m\}_{m=1:M}$. Assume that the total number of measurements is $d$. There are $m_j, m_j<d,$ measurements on each bus $m$, where each bus is considered as a local, spatial region, corresponding to one Local CorrDet detector, $\phi_m$.

\begin{figure}[h]
 \begin{center}
 \includegraphics[height=7cm]{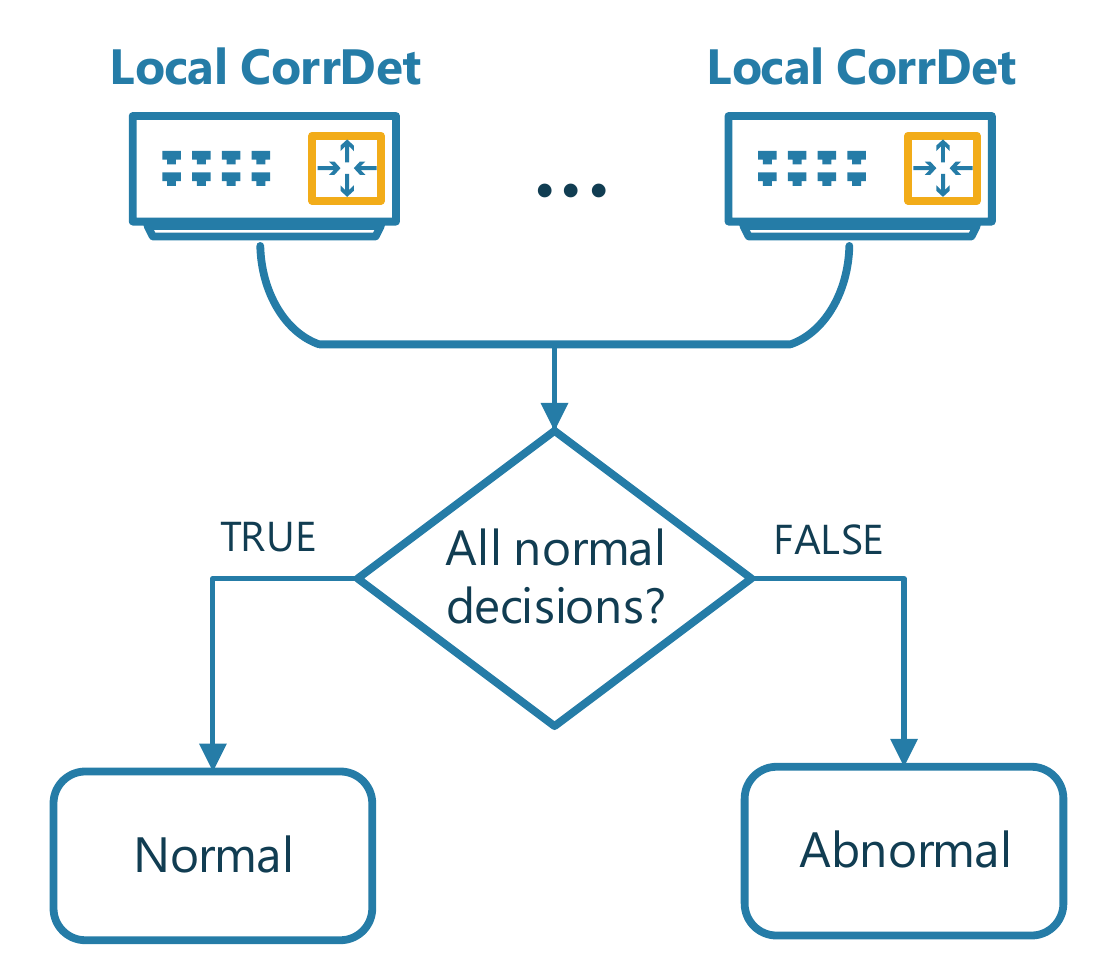}
 \caption{Voting strategy for Ensemble CorrDet detector}
 \label{fig:local}
 \end{center}
 \end{figure}

For $\Phi_R$, the learning process consists of estimating $\mathbf{\mu}$ and $\mathbf{\Sigma}^{-1}$ from normal training samples $\mathbf{z_i}$ ($\mathbf{z_i} \in \mathbb{R}^{1 \times d}$). A similar strategy is proposed to learn the Ensemble CorrDet detector. The learning of $\Phi_E$ involves the estimation of a set of Local CorrDet detectors, $\phi_m$. For each $\phi_m$, similarly, the learning process consists of estimating its $\mathbf{\mu}_m$ and $\mathbf{\Sigma}^{-1}_m$ from the normal training samples with selected measurements $\mathbf{z_{i,m}}$ ($\mathbf{z_{i,m}}$ is a $1 \times m_j$ vector). The threshold value $\tau_m$ for each $\phi_m$ is estimated using the same strategy of CorrDet detector as shown in Eq.\ref{eq:threshold}.  For the new incoming samples, a set of squared Mahalanobis distances, $\Phi_E$, are computed and compared with the corresponding set of thresholds, $T$, where $T = \{\tau_m\}_{m=1:M}$. If at least one squared Mahalanobis distance in $\Phi_E$ is greater than its corresponding threshold, this incoming sample is classified as an anomaly. Otherwise, it is classified as normal samples.  This voting process is visualized in Figure \ref{fig:local}.

Let $K_1$ and $K_2$ the number of training and testing samples, respectively. Let $\mathbf{Z}$ ($\mathbf{Z} \in \mathbb{R}^{d \times K_1}$) and $\mathbf{\widetilde{Z}}$ ($\mathbf{\widetilde{Z}} \in \mathbb{R}^{d \times K_2}$) the training and testing samples, respectively. Let $\mathbf{Y}$ ($\mathbf{Y} \in \mathbb{R}^{1 \times K_1}$) and $\mathbf{\widetilde{Y}}$ ($\mathbf{\widetilde{Y}} \in \mathbb{R}^{1 \times K_2}$) the corresponding labels. $\mathbf{\delta}_{Z,m}$ ($\mathbf{\delta}_{Z,m} \in \mathbb{R}^{1 \times K_1}$) denotes the squared Mahalanobis distances of all training samples with respect to $m$th CorrDet detector, $\phi_m$. $\mathbf{\delta}_{\widetilde{z}_k}$ ($\mathbf{\delta}_{\widetilde{z}_k}\in \mathbb{R}^{1 \times M}$) denotes the squared Mahalanobis distances of $k$th testing sample with respect to all CorrDet detectors, $\Phi_E$.  The pseudo code for the proposed Ensemble CorrDet algorithm is shown in the following. 

\begin{algorithm}
   \caption{Ensemble CorrDet algorithm}
    \begin{algorithmic}[1]
    	\State  \emph{Train an Ensemble CorrDet detector:} 
      \Require $\mathbf{Z}$, $\mathbf{Y}$, $\mathbf{\widetilde{Z}}$
      \For {Every Local CorrDet detector $m = 1:M$} 

      \State Compute the mean and covariance of normal statistics: $\mathbf{\mu}_m$ and $\mathbf{\Sigma}^{-1}_m$
      \State Compute the squared Mahalanobis distance: compute $\mathbf{\delta}_{Z,m}$ using Eq. \ref{CorrDet-algorithm}
      \State Compute the threshold: $\tau_m$
            
      \EndFor
          	\State  \emph{Test using the Ensemble CorrDet detector:}

      \For {Every test sample $k = 1: K_2$}
      
      \State Compute the squared Mahalanobis distance: compute $\mathbf{\delta}_{\widetilde{z}_k}$ using Eq. \ref{CorrDet-algorithm}
      
      \If {$\forall m,  \mathbf{\delta}_{\widetilde{z}_k} < \tau_m$}
      \State Classify $\widetilde{z}_k$ as normal sample: $\widetilde{y}_k = 0$
      \Else
      \State Classify $\widetilde{z}_k$ as abnormal sample: $\widetilde{y}_k = 1$
      \EndIf
      
     \EndFor
      \Ensure $\mathbf{\widetilde{Y}}$
\end{algorithmic}
\end{algorithm}



%

\subsection{Fusion of Physics Model-Based Data Driven Detection Methods} \label{sec:combined}

In order to employ the anomaly detection capabilities of both state estimator solution and the Ensemble CorrDet algorithm, we fuse the results from both the methodologies as shown in Figure \ref{fig:dist-sub}. In the Ensemble CorrDet algorithm, each sample in the testing set will have $M$ CorrDet distances, one for each region as its decision score. We need an overall ECD decision score for each sample to combine it with the decision score from state estimator ($\mathbf{\Psi}_{SE, \widetilde{z}_k}$). SE decision score is calculated as shown in equation \eqref{eq:se_score}.

\begin{equation} \label{eq:se_score}
    \mathbf{\Psi}_{SE} = \sum_{i=1}^d \left[\frac{CME_i}{\sigma_i}\right]^2.
\end{equation}

To find the overall ECD decision score ($ \mathbf{\Psi}_{ECD, \widetilde{z}_k}$) for each testing sample, we consider the squared Mahalanobis distance of Local CorrDet detector which led us to decide whether that sample is anomalous or normal.  This is achieved by considering the decision score of the region whose Local CorrDet detector detected an anomaly. In case that there are multiple Local CorrDet detectors that detect an anomalous sample, the maximum decision score is selected from these multiple Local CorrDet detectors. If there was no anomaly detected from any local CorrDet detectors, the minimum of all decision scores from local CorrDet detectors is considered as the ECD decision score.

We normalize the decision scores obtained from state estimator and Ensemble CorrDet detector by subtracting by their corresponding mean value and dividing by their corresponding standard deviation to form $\mathbf{\Psi}_{ECD_{normalized}}$ and $\mathbf{\Psi}_{SE_{normalized}}$. For each testing sample, we add the normalized decision scores from the state estimator and Ensemble CorrDet detector, and create a new decision score termed as Fusion decision score ($\mathbf{\Psi}_{fusion, \widetilde{z}_k}$). Fusion decision scores, which are calculated as shown in equation \eqref{eq:fusion}, are compared with ground truth values to show the improvement in fused model performance compared to individual detectors.

\begin{equation} \label{eq:fusion}
    \mathbf{\Psi}_{fusion} = \mathbf{\Psi}_{ECD_{normalized}} +  \mathbf{\Psi}_{SE_{normalized}}.
\end{equation}

In order to decide whether an incoming test sample is anomalous or normal, we define a threshold ($\mathbf{\tau}_{fusion}$) based on the fusion decision scores of training set using equation \eqref{eq:threshold}. We compare the fusion decision score of test sample ($\mathbf{\Psi}_{fusion, \widetilde{z}_k}$) with ($\mathbf{\tau}_{fusion}$) as shown in equation \eqref{fusion_decision} to predict whether it is anomalous or normal. 

\begin{equation} \label{fusion_decision}
  \widetilde{z}_{k} =\begin{cases}
    anomalous, & \mathbf{\Psi}_{fusion, \widetilde{z}_k} \geq \mathbf{\tau}_{fusion} \\
    normal, & \mathbf{\Psi}_{fusion, \widetilde{z}_k} < \mathbf{\tau}_{fusion}.
  \end{cases}
\end{equation}

\section{Case Study} \label{sec:case}


\subsection{Dataset description}

\begin{figure*}[t]
\begin{center}
\includegraphics[width=1.9\columnwidth]{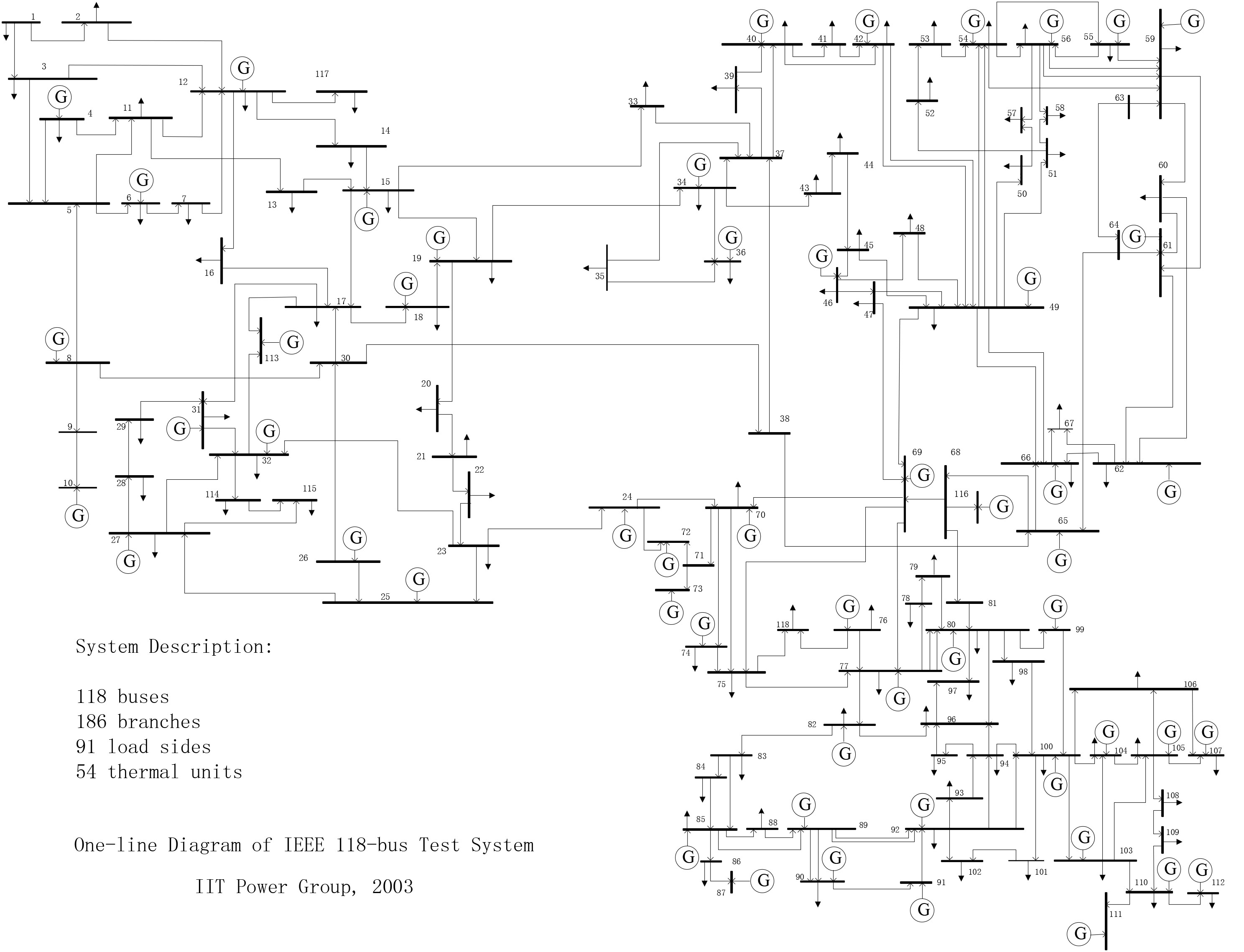}
\caption{IEEE-118 Bus System}
\label{fig:118}
\end{center}
\end{figure*}

For this work, the IEEE 118 bus system \cite{ieee-118} illustrated in Figure \ref{fig:118} was used to generate the training and testing data. Errors are introduced at random in 5\% of the dataset samples and true labels are assigned during this process. Error detection based on physics-based models is performed as a post-processing step to PSSE using CMECST. The dataset consists of 10000 samples with 712 measurements, of which 21 measurements are zero injection measurements. Zero injection measurements are not modeled as equality constraints in the PSSE process. The standard deviation of zero-injection measurements is calculated as shown in (\ref{eq:zero-inj-1}) and (\ref{eq:zero-inj-2}) to avoid any gain matrix singularity issues in PSSE process.  The types of measurements considered were a combination of standard SCADA measurements, i.e. bus voltage magnitudes, real and reactive power injections, and real and reactive power flows. The measurement set data was generated using the power flow options in the MATPOWER package in MATLAB \cite{matpower}. PSSE uses all 712 measurements while the Ensemble CorrDet detector will only use 691 measurements, excluding the zero injection measurements. A drifting load profile was considered for the generation of the measurement set. The drift was modeled by the Ornstein-Uhlenbeck (O-U) process - \textit{a mean-reverting process} \cite{bibbona,thierfelder2015trending}. This is a stochastic process similar to a random walk, but has a tendency to drift back towards the original load. The mean loading condition is updated periodically to model physical reality as aptly as possible. This presents a greater challenge to the data driven solution since there will be greater variations from the mean vector. A detailed discussion of O-U process is included in the subsequent section.

\subsubsection{Ornstein-Uhlenbeck (O-U) Process}
The O-U process $X_{t}$ is defined by the following stochastic differential equation (SDE):
\begin{equation}
\label{eq:o-u-1}
dX_t = -\beta (X_t - \mu_{o-u})dt + \sigma_{n} dW_t 
\end{equation}
where $X_t$ is a random variable,
$W_t$ is the driving noise,
$\beta$ is the decay-rate,
$\sigma^2_{n}$ variance of noise, and 
$\mu_{o-u}$ is the long term mean.

(\ref{eq:o-u-1}) can be solved using Ito's formula and the solution is given by (\ref{eq:ito-sol-1}). It can be seen from (\ref{eq:ito-sol-1}) that $\lim_{t \to \infty} X_{t} = \mu_{o-u} $.

\begin{equation}
\label{eq:ito-sol-1}
X_t  = e^{-\beta t}X_0 + \mu_{o-u}\left(1 - e^{-\beta t}\right) + \sigma_{n}\int_{0}^{t}e^{\beta \left(t_{0} - t\right)} dW_{t_{0}}
\end{equation}

\subsection{Experimental Results}

The anomaly detection problem can be treated as a classification problem, as our goal is to classify testing samples into normal and anomalous classes based on the models trained on training samples. To evaluate the performance of strategies included in this paper, we make use of classification metrics \cite{Bradley1997} such as Receiver Operating Characteristics (ROC) curves and Area Under Curve (AUC) score, which are defined based on True Positive Rate (TPR) and False Positive Rate (FPR) of a classification model. ROC curves represent classification performance of a classifier by varying thresholds and calculating TPR and FPR values for each threshold for decision scores. In our analysis, we plot ROC curves using $\mathbf{\Psi}_{fusion}$ values and ground truth values.

\begin{figure}[h]
\begin{center}
\includegraphics[height=6cm]{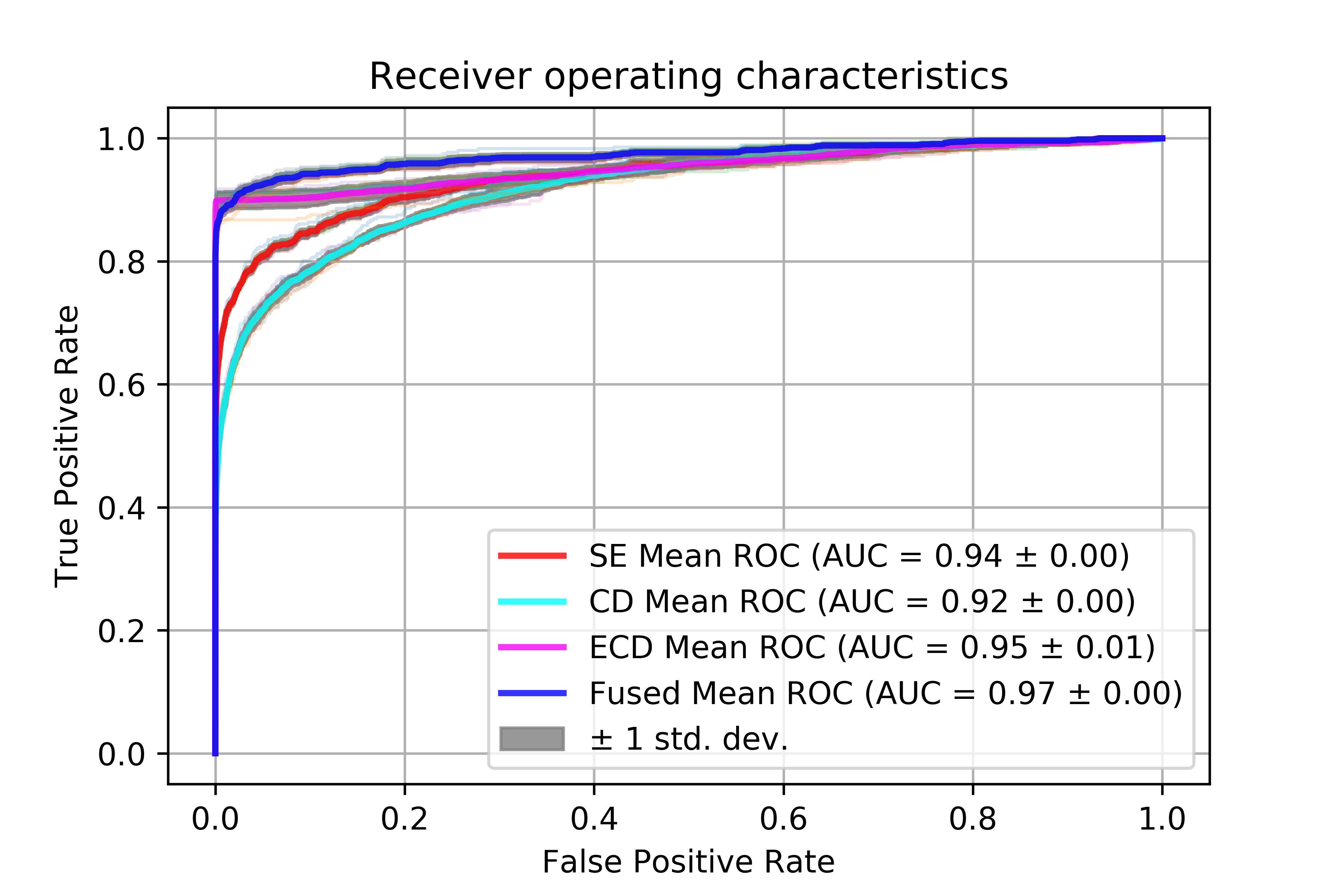}
\caption{ROC curve comparison}
\label{fig:roc}
\end{center}
\end{figure}

In our dataset, we used 30\% of samples for training and 70\% of the samples for testing the model's anomaly detection abilities. In order to reduce the bias of our models, we repeated model training and testing for different combinations of samples in the data for 10 times. For each combination, we obtained a ROC curve for testing samples to understand the model performance. In Figure \ref{fig:roc}, we show the ROC curves for multiple experiments conducted along with a mean ROC curve (averaged over FPR) and variation of mean AUC score (averaged over FPR) for each of the methodologies proposed in this paper.  

In Figure \ref{fig:roc}, we can notice that mean AUC score for Ensemble CorrDet detector (0.95) is better than mean AUC score for state estimator (0.94), but the mean AUC score for fusion results (0.97) is much better than both Ensemble CorrDet algorithm and state estimator method anomaly detection methodologies. The proposed hybrid data-driven physics model-based anomaly detection methodology improved mean AUC score by 3.2\% compared to state estimator results.

\begin{figure}[h]
\begin{center}
\includegraphics[height=6cm]{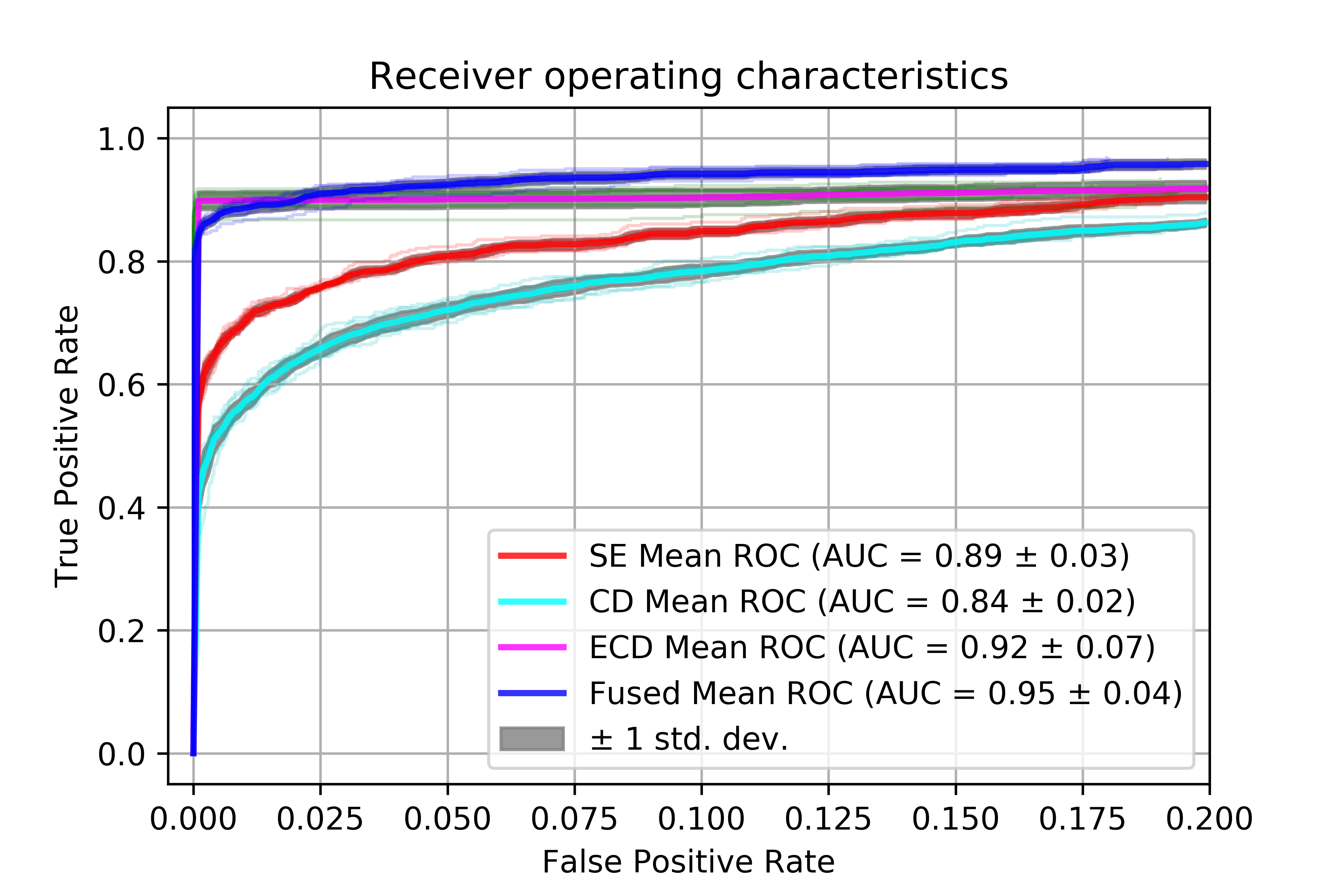}
\caption{ROC curve comparison for a False Positive Rate less than 0.2}
\label{fig:roc_trimmed}
\end{center}
\end{figure}

We can also notice that Ensemble CorrDet detector performs much better compared to state estimator solution and CorrDet detector as shown in Figure \ref{fig:roc_trimmed}. One reason is due to the fact that using all the measurements in CorrDet detector incurs numerical issues but when we look at Ensemble CorrDet detector, we are reducing the measurements for individual detectors, thereby reducing the numerical issues which improves the anomaly detection capabilities.

For an anomaly detection problem, FPR should be fairly low or the model ends up wrongly classifying many normal samples as anomalous. Hence, we also show the performance of the presented anomaly detection methodologies specifically for FPR values less than 0.2 through corresponding mean ROC curves and variation of mean AUC scores for multiple experiments in Figure \ref{fig:roc_trimmed}. The optimal threshold ($\mathbf{\tau}_{fusion}$) for fusion decision scores ($\mathbf{\Psi}_{fusion}$) also lies in this region as to maximize F1-score, FPR value has to be fairly low.

In Figure \ref{fig:roc_trimmed}, instead of absolute AUC score, we show the relative AUC score which is the ratio of AUC score by 0.2 (area in the curve for FPR values between 0 and 0.2). The presented hybrid data-driven physics model-based anomaly detection methodology improved mean AUC score by 6.75\% compared to state estimator results for FPR values than 0.2.

From Figures \ref{fig:roc} and \ref{fig:roc_trimmed}, we can observe that the performance improvement from the presented hybrid data-driven physics model-based anomaly detection methodology compared to state estimator prediction is larger in the ROC space where FPR is less than 0.2 compared to overall ROC space, which is a favorable outcome.


\section{Conclusion} \label{sec:conc}

This paper presents a hybrid physics model-based data-driven framework for the detection of FDI attacks on the SG real-time monitoring.  The physics model-based solution uses the state of the art Innovation Concept of bad data detection and the novel, data-driven Ensemble CorrDet algorithm is introduced to exploit both spatial and temporal characteristics of the SG. The Ensemble CorrDet algorithm uses information from distributed, Local CorrDet detectors to make a decision on whether or not there is an FDI in the system.  By doing this, the data-driven solution not only uses temporal information, but spatial information as well. Decision level fusion is used to combine the information that each individual anomaly detection method contains, considering the confidence that each method has for a given sample.

The hybrid FDI detection framework was tested on the IEEE 118 bus system. Test results show that the fusion of the individual techniques has the best overall performance, detecting FDI attacks at high rate without many false alarms, which can cause issues in a similar fashion to an undetected FDI attack. The presented hybrid framework improved mean Area Under Curve score for testing set by 6.75\% compared to physics model-based results. The fusion of physics model-based data-driven solutions and of temporal and spatial information has been shown to be an improvement on the detection of FDI attacks on the SG, opening up opportunities for future research in this area.

\section*{Acknowledgment}

This material is based upon work supported by the National Science Foundation under Grant Number 1809739.

\bibliographystyle{iet}
\bibliography{sdnreferences,smartgridrefs,smartgrid,data-set,mlrefs}

\end{document}